\documentclass[aps,pra,twocolumn,amssymb,amsfonts,floatfix,showpacs]{revtex4}

\usepackage{graphicx}
\usepackage{dcolumn}
\usepackage{bm}

\begin{document}

\title{Creation of Stable Multipartite Entangled States in Spin Chains
with Defects}

\author{Lea F. Santos}
\email{Lea.F.Dos.Santos@Dartmouth.edu}
\affiliation{\mbox{Department of Physics and Astronomy, 
Dartmouth College, Hanover, NH 03755, USA}}

\begin{abstract}
We show how defects in a spin chain described by the $XXZ$ model 
may be used to generate entangled states, such as
Bell and $W$ states, and how to maintain them
with high fidelity. In the 
presence of several excitations, we also discuss how the anisotropy of 
the system may be combined with defects to 
effectively assist in the creation of the desired states.

\end{abstract}

\maketitle

\section{Introduction}

Applying the superposition principle to product states of 
composite systems leads to one of the most striking 
properties of quantum mechanics: entanglement. What 
used to be a subject within the foundations of quantum
mechanics is now a resource for 
the development of new technologies, playing a central 
role in quantum computation and quantum communication.

The generation of highly entangled states is 
a prerequisite for quantum teleportation, various
quantum cryptographic protocols, and is also needed in quantum computation.
In the bipartite setting, maximal entanglement appears in
the Bell [or also called 
Einstein-Podolsky-Rosen (EPR)] state $(1/\sqrt{2}) 
(|10\rangle + |01\rangle)$
\cite{popescu}. D\"ur {\it et al} \cite{dur} showed that 
there are two different kinds of genuine tripartite
pure state entanglement: the 
maximally entangled Greenberger-Horne-Zeilinger (GHZ) state \cite{GHZ}
and the so called $W$ state. The 
later is the state of three qubits that retains
a maximal amount of bipartite entanglement when any one of the three 
qubits is traced out. It is written as 
$(1/\sqrt{3}) (|100\rangle + |010\rangle + |001\rangle)$.
Here, we show how Bell and $W$ states may be created in 
spin chains with defects.

Spin chains provide ideal systems for the study of entanglement.
They are naturally used to model quantum computers: the two
states of a spin-1/2 particle correspond to the two levels of a qubit
and the exchange interaction corresponds to the qubit-qubit interaction.
We focus here on a spin chain described by the $XXZ$ model, where 
two different kinds of interaction are identified: the Ising part,
proportional to the anisotropy coupling, and the XY part,
which can be used to create entanglement
between two or more qubits \cite{wang,pra}. 

In principle, the qubit level spacings may be controllable
\cite{mark}, allowing the creation of
entanglement between precisely selected qubits. as it turns out, the 
anisotropy of the system can also be used to our advantage,
for states with different numbers of 
neighboring excited qubits are not coupled and 
a large system may then be treated as consisting of several 
uncoupled small chains \cite{pra,pra05}.

The paper is organized as follows. The model is described in the
next section. Section 3 discusses how defects can be used to
create Bell and $W$ states when the system contains
just one excited qubit. Once the desired state is generated,
we outline a way to preserve it with high fidelity for 
a long time, which requires quickly detuning the defects
involved in the process. Section 4 shows how the anisotropy may 
be used as an extra tool toward the creation of entanglement.

\section{The Model}

A spin chain with nearest neighbor coupling is considered. 
The Hamiltonian describing the system has two parts: 
one corresponds to the Zeeman energy of each spin, $H_d$,  and
the other is related to the spin-spin interaction and
is given by the $XXZ$ model, $H_{XXZ}$. In units where 
$\hbar =1$, we have

\begin{equation}
H = H_d + H_{XXZ},
\label{XXZ}
\end{equation}
where

\begin{eqnarray}
&& H_d = \sum _{n=1} ^{L} \frac{\varepsilon _n}{2} \sigma _{n}^{z} ,\nonumber \\
&& H_{XXZ}=\sum _{n=1} ^{L} \left[ \frac{J\Delta}{4}  
\sigma _{n}^{z} \sigma _{n+1}^{z} +  H_{\rm{hop}} 
\right],  \nonumber \\
&&H_{\rm{hop}} = \frac{J}{8} \left(\sigma _{n}^{+} 
\sigma _{n+1}^{-} + \sigma _{n}^{-} \sigma _{n+1}^{+}  \right). \nonumber
\end{eqnarray}
Above $\sigma ^{z,+,-}$ are Pauli matrices.
There are $L$ sites and we deal with a periodic (or closed) chain, that is,
sites $n+L$ and $n$ are the same. Each site $n$ is subjected to a magnetic 
field in the $z$ direction, giving the energy splitting $\varepsilon _n$. In 
terms of qubits, $\varepsilon _n$ is the level spacing of qubit $n$.
A spin pointing up corresponds to an excited qubit, or an excitation.
The parameter $J$ is the hopping integral and $\Delta $ is 
a dimensionless parameter related to the anisotropy 
coupling. The non-diagonal term $H_{\rm{hop}}$ 
is responsible for propagating the excitations and delocalizing the states.
The diagonal term $\sigma _{n}^{z} \sigma _{n+1}^{z}$ 
gives the Ising interaction. It
is only relevant when at least two excitations are present, being therefore
associated with many-body problems. We set 
$J$ and $\Delta>0$.

We note that some recent works have discussed entanglement in 
anisotropic systems with impurities \cite{osterloch,osenda,huang}, but
their anisotropy differs from the one in Hamiltonian (\ref{XXZ}). 
The hopping part of the Hamiltonian considered here 
can be equivalently written as 
$H_{\rm hop}\propto 
J_x \sigma_n^x \sigma_{n+1}^x + J_y \sigma_n^y \sigma_{n+1}^y$. 
Here $J_x=J_y=J$, but in the models
cited above, the degree of anisotropy comes from the difference
between $J_x$ and $J_y$. The anisotropy in our case originates from 
the extra Ising interaction. In terms of entanglement, few studies
have been developed with this model \cite{rigolinIJQI,russian}, though
this is the model that best describes some proposals of quantum 
computers \cite{mark}.

In a homogeneous chain, all qubits have the same 
level spacing $\varepsilon $, while a disordered system has at 
least one site with a different energy $\varepsilon_n = \varepsilon + d_n$,
which is called a defect. In principle, complete control of the qubit level
spacings is available, which allows the `creation of defects'. We 
set $d_n>0$ and 
assume that $\varepsilon $ largely exceeds $J, \Delta$ and $d_n$, 
so the ground state of the system corresponds to all spins pointing down.
In what follows, we count energy off the ground state energy
${\cal E}_{0} = -\sum _{n=1} ^{L} \varepsilon _n/2 + L J\Delta/4$, i.e., we 
replace in Eq.~(\ref{XXZ}) $H\rightarrow H-E_0 $. 

The states $|\alpha_1 \alpha_2 ... \alpha_L\rangle$, where
$\alpha _i$ = 0 or 1, correspond to the computational basis.
To address the different states of the system we use a notation that is
common when determining the energy spectrum of
spin chains with the Bethe ansatz \cite{bethe}.
A state with a single excitation on site $n$,
that is  $|\downarrow _{1} \downarrow _{2}...\downarrow _{n-1} 
\uparrow _{n} \downarrow _{n+1}... \downarrow _{L}\rangle $,
or equivalently $|0_{1} 0_{2}... 0_{n-1}1_{n}0_{n+1}... 0_{L}\rangle $, is
simply written as $\phi (n)$. 
A state with two excitations, one on 
site $n$ and the other on site $m$, is $\phi (n,m)$, 
which is a simplified notation for  
$|\downarrow _{1} \downarrow _{2}...\uparrow _{n} \downarrow _{n+1}...
\uparrow _{m} ... \downarrow _{L}\rangle $, or equivalently
$|0_{1} 0_{2} ...1_{n} 0_{n+1}... 1_{m} ... 0_{L} \rangle $.  

In the system considered here, the $z$ component of the total
spin, $S_z=\sum_{n=1}^{L} S_{n}^{z}$, is conserved, hence states with different
number of excitations are not coupled and the Hamiltonian 
effectively describes uncoupled subsystems. 
The diagonalization of a certain block corresponding 
to $N$ excitations leads to 
$L!/[N!(L-N)!]$ eigenstates, such that the $k$th 
eigenstate is written as 

\begin{eqnarray}
&&|\psi ^{(k )}_{N,L}\rangle = \nonumber \\
&&\sum_{n_1<n_2<... <n_N=1}^L a^{(k)}(n_1,n_2,..., n_N) 
|\phi (n_1,n_2,..., n_N)\rangle . \nonumber
\end{eqnarray}

\section{Single excitation}

Suppose that there is only one excitation in the chain and assume that all
qubits have the same level spacing $\varepsilon $, except for some defects
that have all the same level spacing  $\varepsilon +d $. By choosing $d$
much larger than the hopping integral $J$, we guarantee that states with 
one excitation on a defect do not couple with states that have 
the excitation on the other sites. In terms of energy spectra,
we form two well separated
bands. It is then possible to treat the spin chain as two smaller
and decoupled chains.

\subsection{Two defects: Bell states}

Let us first examine the case where there are simply two defects
with $d\gg J$, placed on sites $n_1$ and $n_2$.
The chain is broken into two, one with $L-2$ qubits,
whose state energies lie within the band ${\cal E}_1 \pm J$, where
${\cal E}_1 = \varepsilon - J\Delta $,
and the other with just two qubits, whose eigenstates
are Bell states and have energies $\sim {\cal E}_1 + d$.
An excitation created on one defect will hop between the two defects
with a frequency determined by their distance. To find this value, we need
to compute the energies $E_{\pm}$ of the two eigenstates
$\psi _{\pm} = \frac{1}{\sqrt{2}} [\phi (n_1) \pm \phi (n_2)]$ 
of the defect chain.

If the two defects are nearest neighbors, $n_1$ and $n_1 +1$, 
they are coupled to first 
order of perturbation theory. We diagonalize a 2x2 submatrix whose diagonal 
elements are ${\cal E}_1 + d$ and off-diagonal elements give 
the effective hopping
integral $J_{\rm eff} = J/2$, so
$E_{\pm} = {\cal{E}}_1 + d \pm J/2$. If the defects are next-nearest neighbors,
$n_1$ and $n_1 +2$,
they are coupled to second order of perturbation theory. Now
the diagonal elements are ${\cal E}_1 + d +J^2/(2d)$ and the off-diagonal 
elements are $J_{\rm eff}=J^2/(4 d)$, leading to states with 
much closer energies
$E_{\pm} = {\cal{E}}_1 + d + J^2/(2d) \pm J^2/(4d) $. The larger the separation
between the defects, 
the closer will be the energies of the two eigenstates.

An excitation initially prepared on site $n_1$ will have probability

\begin{equation}
P_{\phi(n_1)} (t) = \frac{1 + \cos [(E_{+} - E_{-})t]}{2},
\label{T1}
\end{equation}
to be later found on the same site. The oscillation period 
of the excitation
between the defects is inversely proportional to the energy difference
of the two eigenstates. It therefore depends on the number 
$\mu $ of sites between the defects as $T_{\mu} = T_{0} (2d/J)^{\mu }$, 
where $T_{0} = 2\pi /J$.

Bell states are created when $P_{\phi(n_1)}=P_{\phi(n_2)}=1/2$, that is, 
at every 
instant $t_B=\pi k/[2(E_+ - E_-)]$, where $k$ is an odd number. 
The more separated are the defects the longer is the wait for a maximally 
entangled state to be created, but in principle we can entangle even remote qubits.

\subsubsection{Detuning procedure}

\begin{figure}[th]
\includegraphics[width=2.5in]{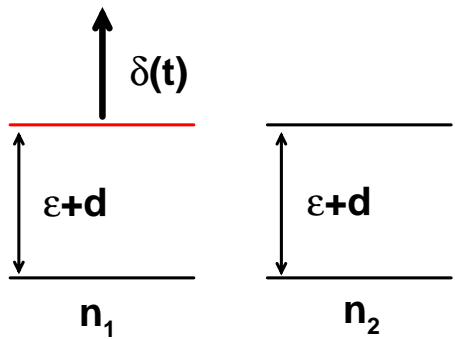}
\caption{Scheme for the detuning procedure for two
defects on sites $n_1$ and $n_2$.}
\label{scheme}
\end{figure}

Once the desired state has been generated, the two 
defects should be detuned for the
state to be maintained unchanged. 
A scheme for a possible detuning procedure is shown in Fig.~\ref{scheme}.
The level spacing of one qubit is kept constant, while the other 
increases according to a certain function of time $\delta (t)$, 
which depends on the experimental 
setting and on the fidelity we want to attain. At the end of the process, 
the energy difference between the two defects should be much larger
than their effective hopping integral.

The detuning should start at $t=t_B$ and, apart from constants, it
is well represented by the Hamiltonian

\begin{equation}
H_{\rm detun}=\delta (t)\frac{1+\sigma_{n_1}^z }{2}
+\frac{J_{\rm eff} }{4} \left(\sigma _{n_1}^{+} 
\sigma _{n_2}^{-} + \sigma _{n_2}^{+} \sigma _{n_1}^{-}  \right). 
\end{equation}

\begin{figure}[th]
\includegraphics[width=3.2in]{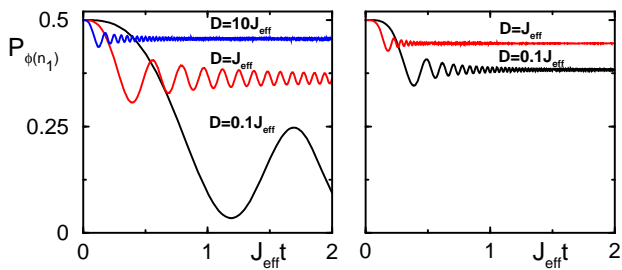} 
\caption{Time evolution of the probability to 
find the excitation on site $n_1$ during the detuning of two defects placed
on $n_1$ and $n_1 +2$. The left panel has $\delta (t)=Dt$ and the right 
panel has $\delta (t)=Dt^2$.
The detuning is initiated at the moment a Bell state is created. 
The effective hopping integral is $J_{\rm eff}=J^2/(4 d)$,
$d=10 J$, and the values of $D$ are shown in the figure.}
\label{detun1}
\end{figure}

In Fig.~\ref{detun1} we show the results of a linear detuning,
$\delta (t)=Dt$, and a quadratic one, $\delta (t)=Dt^2$. Two
defects with $d=10J$ coupled to second order of perturbation theory
are considered. The larger
the ratio $D/J_{\rm eff}$, the closer the generated state will be to the
Bell state at later times. Contrary to what happens when creating states, 
now more separated defects make it easier to preserve the state with
high fidelity. 

\subsubsection{Measures of entanglement}

Several measures of entanglement have been introduced. Here we briefly
comment on how two of them, concurrence \cite{wootters} and global entanglement
\cite{jpa,viola,brennen}, can be affected by defects. They both vary from 0 
(unentangled state) to 1 (entangled state), but concurrence
is a measure of bipartite entanglement, while the other is able to
quantify multipartite entanglement. 

It is straightforward to conceive a system where all eigenvectors
are Bell states and therefore have maximum individual
(as well as the average over all states) concurrence.
An example is a chain with an even number of qubits $L$, where
pairs of sites have the same level spacings, but differ from 
the others by more than the coupling strength between them. 
In this situation, all eigenvectors are Bell states 
each involving a particular pair of 
resonant sites. In terms of multipartite entanglement, designing
a distribution of defects that would lead to 
large global entanglement is in general more demanding.
The definition of global entanglement is 
$2-(2/L)\sum_{n=1}^L {\rm tr}\rho_{n}^2$, where $\rho_n$ is the
reduced density matrix for qubit $n$ \cite{jpa,viola}.
For the example described, it
is maximum only for a two-qubit chain, decreasing with 
the chain size as $2 - 2/L - 2(L-2)/L$. 
Moreover, systems described by the $XXZ$ model may not be ideal, because
states that are known to have large global entanglement,
such as the $GHZ$ or products of Bell states, cannot be created. This is
a consequence of the conservation of the total spin in the $z$ direction,
which prevents the coupling between basis states that have a different 
number of excitations.

\subsection{Three defects: $W$ states}

To create a $W$ state with one excitation, three defects are necessary.
The procedure is very similar to the one described before. Let us assume
three nearest neighbor defects, $n_1$, $n_2=n_1+1$, and $n_3=n_1+2$,
with $d \gg J$. The three eigenvectors of the defect
chain are a good approximation to the actual corresponding vectors 
obtained with 
the total Hamiltonian. 
They can be simply computed by diagonalizing the submatrix

\[
\left(
\begin{array}{ccc}
{\cal E}_1 +d & J/2 &0 \\
J/2 & {\cal E}_1 +d & J/2 \\
0 & J/2 & {\cal E}_1 +d \end{array}
\right) ,
\]
giving

\[
\begin{array}{lll}
\psi_{a} = \frac{1}{2} [\phi(n_1)+\sqrt{2}\phi(n_2)+\phi(n_3)] 
& \hspace{0.3 cm} & E_a={\cal E}_1 +d+J/\sqrt{2},  \\
\psi_{b} = \frac{1}{\sqrt{2}} [-\phi(n_1)+\phi(n_3)]  
& \hspace{0.3 cm} & E_b={\cal E}_1 +d , \\
\psi_{c} = \frac{1}{2} [\phi(n_1)-\sqrt{2}\phi(n_2)+\phi(n_3)]  
& \hspace{0.3 cm} & E_c={\cal E}_1 +d -J/\sqrt{2}. \end{array}
\]

The $W$ state is generated by preparing an initial state with an excitation
in the middle site $n_2$, since this defect is coupled to the other 
two to first
order of perturbation theory. The probability for the excitation to be found
on this same site later in time is

\begin{equation}
P_{\phi(n_2)} (t) = \frac{1 + \cos [(E_{a} - E_{c})t]}{2},
\end{equation}
while the probabilities for sites $n_1$ and $n_3$ are the same
\begin{equation}
P_{\phi(n_1)} (t)=P_{\phi(n_3)} (t) = \frac{1 - \cos [(E_{a} - E_{c})t]}{4}.
\end{equation}

For the $W$ state we need 
$P_{\phi(n_1)} (t)=P_{\phi(n_2)} (t)=P_{\phi(n_3)} (t)=1/3$, which happens
at every instant 

\begin{equation}
t_W=\frac{(-1)^k \arccos (-1/3) + 
2 \pi (k - \lfloor k/2 \rfloor )}{E_{a} - E_{c}},
\end{equation}
where $k$ is an integer.

\begin{figure}[htb]
\includegraphics[width=3.2in]{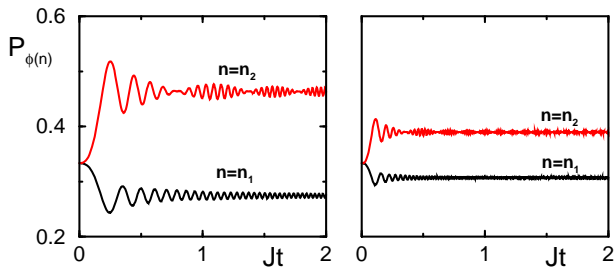} 
\caption{Time evolution of the probability to 
find the excitation on site $n_1$ and on site $n_1+1$ 
during the detuning of three neighbor defects. The probability 
for the third defect placed on $n_1+2$ is very close to
the probability for $n_1$. 
The detuning is initiated at the moment a $W$ state is created. For
the left panel $D_1=10J$ and $D_2=100J$, for the right panel,
$D_1=50J$ and $D_2=500J$.}
\label{fig2}
\end{figure}

A possible detuning procedure for the defects for $t\geq t_W$ 
is the following: the level spacing of defect $n_3$ is kept fixed,
while the level spacings of $n_1$ and $n_2$ increase in time
according to the functions $\delta _1(t)$ and
$\delta _2 (t)$, respectively.
The Hamiltonian describing the process is given by

\begin{eqnarray}
H_{\rm detun}&=&\delta_1 (t) \frac{1 +\sigma_{n_1}^z }{2} +
\delta_2 (t) \frac{1 +\sigma_{n_2}^z }{2} \nonumber \\
&+&\frac{J }{2} \sum_{n=n_1,n_2}\left(\sigma _{n}^{+} 
\sigma _{n+1}^{-} + \sigma _{n+1}^{+} \sigma _{n}^{-}  \right). 
\end{eqnarray}
For the three defects to become out of resonance, we can use
for instance, $\delta_1(t)=D_1 t$, $\delta_2(t)=D_2 t$,
and $D_1\neq D_2$.
The faster the detuning is performed the higher the fidelity for the
desired state. Moreover, the final 
result becomes better if $D_2>D_1$ instead of $D_1>D_2$. 
An example is shown in Fig~\ref{fig2}.

\section{Multiple excitations}

\begin{figure}[htb]
\includegraphics[width=2.8in]{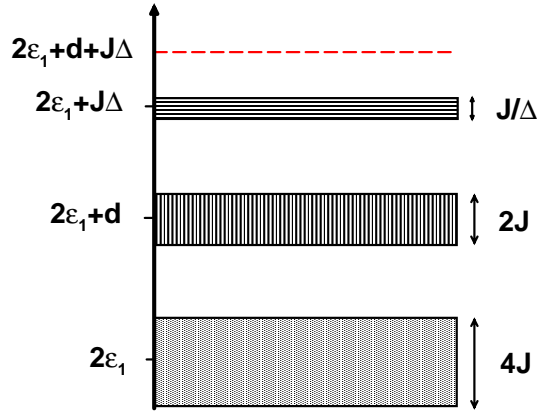}
\caption{Energy spectrum of the considered spin chain with two excitations
and one defect with $J\Delta \gg d \gg J$.}
\label{diagram}
\end{figure}

When more than one excitation is present in the chain, 
the Ising part of the Hamiltonian
starts having an effect. If the anisotropy parameter $\Delta $ is large,
i.e. $J\Delta\gg J$, several well separated energy bands appear, each
corresponding to states with a different number of excitations 
next to each other.
The case of two excitations, for instance, is schematically shown in 
Fig.~\ref{diagram}. The 
states where the excitations are far from each other have 
energy inside the band $2{\cal E}_1 \pm 2J$, while the states
where they are next to each other have a much higher energy
and form a much narrower band $2{\cal E}_1 + J\Delta +J/(2 \Delta )
\pm J/(2 \Delta )$. Such bound pairs are coupled to second
order of perturbation theory, taking much longer 
to propagate through the chain if compared to the free excitations.
The two bands are not coupled and can be treated separately.

In addition, suppose that
a single defect $n_1$ is added to the chain with two
excitations, such that $J\Delta \gg d\gg J$. Two new bands appear
[see Fig.~\ref{diagram}]. One band has states with
the two excitations far from each other, but one
trapped on the defect, they have energy 
$2{\cal E}_1 + d\pm 2J$. The other band is
made of just two (Bell) states having the largest energies. Here,
the two excitations are next to each other and 
one of them is on the defect site. We have,

\begin{eqnarray}
&&\psi _{\pm} = \frac{1}{\sqrt{2}} [\phi(n_1-1,n_1) \pm \phi (n_1,n_1+1)]
\\
&&E_{\pm} = 2{\cal{E}}_1 +d +J\Delta + \frac{J}{4\Delta} +
\frac{J^2}{4(J\Delta +d)} \pm \frac{J^2}{4(J\Delta +d)}. \nonumber
\end{eqnarray}  

Advantage may be taken of the large anisotropy to create entanglement.
Preparing an initial bound pair on the sites $(n_1-1,n_1)$, 
for example, will lead to Bell states at every instant
$t_{BP} = 2(J\Delta +d) [\pi/2 + k \pi]/J^2$, where $k$ is odd.
As before, detuning site $n_1-1$ or site $n_1+1$ should
preserve the maximally entangled state. Moreover,
since states $(n_1-1,n_1)$ and $(n_1,n_1+1)$
are coupled to second order, 
the interaction strength is small and the detuning 
rate required to keep the created Bell state with high fidelity
does not need to be very large.
Several other possibilities for Bell and $W$ states where
use is made of the defects and anisotropy have been
discussed elsewhere \cite{pra,pra05}.

\section{Conclusion}

We have investigated how bipartite and tripartite states can be generated in 
strongly anisotropic spin-1/2 chains described by the 
$XXZ$ model by controlling the Zeeman energy
of the spins [or equivalently, by manipulating the qubit level spacings].
Even though the interaction between the qubits is always on, 
it is effective only between states with very close energy.
Selected resonant defects can then be used
in the creation of entangled states.
Once generated, the maintenance of the desired states 
depends on how fast the defects involved
in the process can be moved away from resonance.

Our interest in the $XXZ$ model emerges from its relevance to
some proposals for quantum computers \cite{mark}, but 
we now intend to extend the method developed here
to other models, especially
those that they may be used to create multipartite states with large
global entanglement.

\begin{acknowledgments}
We acknowledge support from Constance and Walter Burke through their
Special Projects Fund in Quantum Information Science, and 
from the organizing committee of the Workshop `Quantum entanglement
in physical and information sciences' (Pisa December 14-18, 2004).
We also thank Mark I. Dykman for introducing to us the idea 
of controlling qubit level spacings and its 
usefulness for quantum gate operations.
We are especially grateful to  Carlos O. Escobar, Gustavo Rigolin 
and Lorenza Viola for a critical
reading of the manuscript and for constructive suggestions.
\end{acknowledgments}


\begin{thebibliography}{0}

\bibitem{popescu} C. H. Bennett, H. J. 
Bernstein, and S. Popescu, Phys. Rev. A {\bf 53}, 2046 (1996).

\bibitem{dur} W. D\"ur, G. Vidal, and J. I. Cirac, Phys. Rev.
A {\bf 62}, 062314 (2000). 

\bibitem{GHZ} D. M. Greenberger, M. Horne, and A. Zeilinger, 
{\it Bell's Theorem, Quantum Theory, and Conceptions of the Universe},
edited by M. Kafatos (Kluwer, Dordercht, 1989).

\bibitem{wang} X. Wang, {\it Phys. Rev.} A {\bf 64}, 012313 (2001).

\bibitem{pra} L. F. Santos, {\it Phys. Rev.} A {\bf 67}, 062306 (2003).

\bibitem{mark}  M. I. Dykman and P. M. Platzman, {\it Fortschr. Phys} 
{\bf 48}, 9 (2000); P. M. Platzman and M. I. Dykman, {\it Science} {\bf 284},
1967 (1999); 

\bibitem{pra05} L. F. Santos and G. Rigolin, {\it Phys. Rev.} A 
{\bf 71}, 032321 (2005).

\bibitem{osterloch} A. Osterloh, L. Amico, G. Falci, and R. Fazio, 
{\it Nature} (London) {\bf 416}, 608 (2002).

\bibitem{osenda} O. Osenda, Z. Huang, and S. Kais, {\it Phys. Rev.} 
A {\bf 67}, 062321 (2003).

\bibitem{huang} Z. Huang, O. Osenda, and S. Kais, {\it Phys. Lett.} 
A {\bf 322}, 137 (2004).

\bibitem{rigolinIJQI} G. Rigolin, {\it Int. J. Quant. Inf.} {\bf 2}, 393 (2004).

\bibitem{russian} P. F. Kartsev and V. A. Kashurnikov, {\it JETP Letters} 
{\bf 80}, 44 (2004).

\bibitem{bethe} H. A. Bethe, Z. Phys. {\bf 71}, 205 (1931);
C. N. Yang and C. P. Yang, Phys. Rev. {\bf 150}, 321, 327 (1966);
M. Karbach and G. M\"uller, e-print cond-mat/9809162;
F. C. Alcaraz, M. N. Barber, and M. T. Batchelor,
Ann. of Phys. {\bf 182}, 280 (1988).

\bibitem{wootters} S. Hill and W. K. Wootters, Phys. Rev. Lett. {\bf 78}, 5022
(1997);
W.K. Wootters, Phys. Rev. Lett. {\bf 80}, 2245 (1998).

\bibitem{jpa} D. A. Meyer and N. R. Wallach, J. Math. Phys. (N.Y.)
{\bf 43}, 4273 (2002).

\bibitem{viola} H. Barnum, E. Knill, G. Ortis, R. Somma, and
L. Viola, {\it Phys. Rev. Lett.}, {\bf 92}, 107902 (2004).

\bibitem{brennen} G. K. Brennen and S. S. Bullock, 
Phys. Rev. A {\bf 70}, 052303 (2004). 

\end{thebibliography}
\end{document}